\documentclass[12pt]{iopart}

\usepackage{color}

\usepackage{bm}
\usepackage{amssymb}
\usepackage{graphicx}
\usepackage{array}
\usepackage{tabulary}
\usepackage{caption}

\usepackage{etoolbox}
\apptocmd{\sloppy}{\hbadness 10000\relax}{}{}

\newcolumntype{K}[1]{>{\centering\arraybackslash}p{#1}}

\begin{document}

\title{Hidden-Beauty Broad Resonance $Y_b(10890)$ in Thermal QCD}

\author{J.Y.~S\"ung\"u $^{1}$,~A. T\"{u}rkan$^{2}$,~H. Da\u{g}$^{2,3}$ and E. Veli Veliev $^{1,4}$}

\address{$^1$ Department of Physics, Kocaeli University, 41380
Izmit, Turkey}

\address{$^2$ \"{O}zye\u{g}in University, Department of Natural and
Mathematical Sciences, \c{C}ekmek\"{o}y, Istanbul, Turkey}

\address{$^3$ Physik Department, Technische Universit\"{a}t M\"{u}nchen, D-85747
Garching, Germany}

\address{$^4$ Education Faculty, Kocaeli University,
41380 Izmit, Turkey}

\begin{abstract}
In this work, the mass and  pole residue of resonance $Y_b$ is
studied by using QCD sum rules approach at finite temperature.
Resonance $Y_b$ is described by a diquark-antidiquark tetraquark current,
and contributions to operator product expansion are calculated by
including QCD condensates up to dimension six. Temperature dependences of the
mass $m_{Y_b}$ and the pole residue $\lambda_{Y_b}$ are investigated. It is seen that near a critical
temperature $(T_c\simeq190~\mathrm{MeV})$, the values of $m_{Y_b}$
and $\lambda_{Y_b}$ are decreased to $87\%$, and to $44\%$ of
their values at vacuum.
\end{abstract}

\section{Introduction}

Heavy quarkonia systems provide a unique laboratory to search the
interplay between perturbative and nonperturbative effects of QCD.
They are non-relativistic systems in which low energy QCD can be
investigated via their energy levels, widths, and transition
amplitudes \cite{Segovia:2016xqb}. Among these heavy quarkonia
states, vector charmonium and bottomonium sectors are
experimentally studied very well, since they can be detected
directly in $e^+e^-$ annihilations. In the past decade,
observation of a large number of bottomonium-like states in
several experiments increased the interest in these structures
\cite{Abe:2007tk,Aubert:2008ab,Huang:2006em,Chen:2008xia,Bonvicini:2005ci}.
However, these observed states could not be conveniently explained
by the simple $q\bar{q}$ picture of mesons. The presumption of
hadrons containing quarks more than the standard quark content
($q\bar{q}$ or $qqq$) are introduced by a perceptible model for
diquarks plus antidiquarks, which was developed by Jaffe in 1976
\cite{Jaffe:1976yi}. Later Maiani, Polosa and their collaborators
proposed that the X, Y, Z mesons are tetraquark systems, in which
the diquark-antidiquark pairs are bound together by the QCD color
forces \cite{Maiani:2004vq}. In this color configuration, diquarks
can play a fundamental role in hadron spectroscopy. Thus, probing
the multiquark matter has been an intensely intriguing research
topic in the past twenty years and it may provide significant
clues to understand the non-perturbative behavior of QCD.

In 2007, Belle reported the first evidence of
$e^+e^-\rightarrow\Upsilon(1S)\pi^+\pi^-, \Upsilon(2S)\pi^+\pi^-$
and first observation for
$e^+e^-\rightarrow\Upsilon(3S)\pi^+\pi^-, \Upsilon(1S)K^+K^-$
decays near the peak of the $\Upsilon(5S)$ state at
$\sqrt{s}=10.87~\mathrm{GeV}$ \cite{Abe:2007tk}. Assigning these
signals to $\Upsilon(5S)$, the partial widths of decays
$\Upsilon(5S)\rightarrow\Upsilon(1S)\pi^+\pi^-$ and
$\Upsilon(5S)\rightarrow\Upsilon(2S)\pi^+\pi^-$ were measured
unusually larger (more than two orders of magnitude) than formerly
measured decay widths of $\Upsilon(nS)$ states. Following these
unusually large partial width measurement, Belle measured the
cross sections of $e^+e^-\rightarrow\Upsilon(1S)\pi^+\pi^-,
\Upsilon(2S)\pi^+\pi^-$ and $\Upsilon(1S)\pi^+\pi^-$, and reported
that the resonance observed via these decays does not agree with
conventional $\Upsilon(5S)$ line shape. These observations led to
the proposal of existence of new exotic hidden-beauty state
analogous to broad $Y(4260)$ resonance in the charmonium sector,
which is a Breit-Wigner shaped resonance with mass
($10888.4^{+2.7}_{-2.6}\pm1.2$) MeV/c$^2$, and width
($30.7^{+8.3}_{-7.0}\pm3.1$) MeV/c$^2$, and is called $Y_b(10890)$
\cite{Chen:2008xia}. In literature, there are several approaches
to investigate the structure of exotic $Y_b$ resonance. In
Ref.~\cite{Chen:2011cta}, $Y_b$ is considered as a
$\Lambda_b\bar{\Lambda}_b$ bound state with an highly large
binding energy. In Refs.~\cite{Albuquerque:2011ix,Zhang:2010mv},
$Y_b$ is interpreted as a tetraquark, and its mass is estimated by
using QCD sum rules at vacuum.

Moreover, it is likely that at very high temperatures within the
first microseconds following the Big Bang, quarks and gluons
existed freely in a homogenous medium called the quark-gluon
plasma (QGP). In 2011, CMS Collaboration reported that charmonium
states $\phi(2S)$ and $J/\psi$ melt or be suppressed due to
interacting with the hot nuclear matter created in heavy-ion
interactions \cite{Chatrchyan:2012lxa,Sirunyan:2016znt}. Following
these observations in the charmonium sector, CMS Collaboration
also reported suppression of bottomonium states, $\Upsilon(2S)$
and $\Upsilon(3S)$ relative to the $\Upsilon(1S)$ ground state
\cite{Sirunyan:2017lzi,Sirunyan:2018nsz}. The dissociation
temperatures for the $\Upsilon$ states are expected to be related
with their binding energies, and are predicted to be $2T_c,~
1.2T_c$ and $T_c$ for the $\Upsilon(1S), \Upsilon(2S)$, $2T_c$,
and $Y(3S)$ mesons, respectively, where $T_c$ is the critical
temperature for deconfinement
\cite{Sirunyan:2018nsz,Mocsy:2007jz,Burnier:2015tda}. Inspiring by
these findings and motivated by the aforementioned discussions, we
focus on the $Y_b$ resonance and its thermal behavior.

This paper is organized as follows. In section\ \ref{sec:Theory},
theoretical framework of Thermal QCD sum rules (TQCDSR) and its
application to $Y_b$ are presented, and obtained analytical
expressions of the mass and pole residue of $Y_b$ are given up to
dimension six operators. Numerical analysis is performed and
results are obtained in section \ref{sec:NumAnal}. Concluding
remarks are discussed in section \ref{sec:Result}. The explicit
forms of the spectral densities are written in Appendix.

\section{Finite Temperature Sum Rules for Tetraquark Assignment}\label{sec:Theory}

QCD sum rules (QCDSR) approach is based on Wilson's operator
product expansion (OPE) which was adapted by Shifman, Vainshtein
and Zakharov, and applied with remarkable success to estimate a
large variety of properties of all low-lying hadronic states
\cite{Shifman,Reinders,Colangelo:2000dp,Narison:2002pw}. Later,
this model is extended to its thermal version that is firstly
proposed by Bochkarev and Shaposnikov, and led to many successful
applications in QCD
\cite{Bochkarev:1985ex,Hatsuda:1992bv,Alam:1999sc,Rapp:2009yu,Mallik:1997kj,Dominguez:2013fca,Veliev:2017fpa}.
In this section, the mass and pole residue of the exotic $Y_b$
resonance are studied by interpreting it as a bound
$[bs][\bar{b}\bar{s}]$ tetraquark via TQCDSR technique which
starts with the two point correlation function
\begin{equation}\label{CorrFunc}
\Pi_{\mu \nu }(q,T)=i\int d^{4}x~e^{iq\cdot x}\langle
\Psi|\mathcal{T}\{\eta_{\mu}(x)\eta_{\nu}^{\dagger}(0)\}|\Psi\rangle,
\end{equation}
where $\Psi$ represents the hot medium state, $\eta_{\mu}(x)$ is
the interpolating current of the $Y_b$ state and $\mathcal{T}$
denotes the time ordered product. The thermal average of any
operator $\hat{O}$ in thermal equilibrium is given as
\begin{equation} \label{operator}
\langle \hat{O}\rangle=\frac{Tr(e^{-\beta \mathcal{H}}\hat{O})}{Tr(e^{-\beta \mathcal{H}})}, \\
\end{equation}
where $\mathcal{H}$ is the QCD Hamiltonian, and
$\mathcal{\beta}=1/T$ is inverse of the temperature, and $T$ is
the temperature of the heat bath. Chosen current $\eta_{\mu}(x)$
must contain all the information of the related meson, like
quantum numbers, quark contents and so on. In the
diquark-antidiquark picture, tetraquark current interpreting $Y_b$
can be chosen as~\cite{Matheus:2006xi}
\begin{eqnarray}\label{TetraCurrent}
\eta_{\mu }(x)&=&\frac{i\epsilon \tilde{\epsilon}}{\sqrt{2}}\Big\{
\Big[~s_{a}^{T}(x)C\gamma_{5} b_{b}(x)\Big] \Big[~
\overline{s}_{d}(x)\gamma_{\mu}\gamma_{5}C\overline{b}_{e}^{T}(x)\Big]  \nonumber \\
&+&\Big[~s_{a}^{T}(x)C\gamma_{5}\gamma_{\mu}b_{b}(x)\Big] \Big[~
\overline{s}_{d}(x)\gamma_{5}C\overline{b}_{e}^{T}(x)\Big]\Big\},
\end{eqnarray}
where $C$ is the charge conjugation matrix and $a,b,c,d,e$ are
color indices. Shorthand notations $\epsilon =\epsilon _{abc}$ and
$\tilde{\epsilon}=\epsilon _{dec}$ are also employed in
Eq.~(\ref{TetraCurrent}).

In TQCDSR, the correlation function given in Eq.~(\ref{CorrFunc})
is calculated twice, as in QCD sum rules at vacuum, in two
different regions corresponding two perspectives; namely the
physical side (or phenomenological side) and the QCD side (or OPE
side). By equating these two approaches, the sum rules for the
hadronic properties of the exotic state under investigation are
achieved. To derive mass and pole residue via TQCDSR, the
correlation function  is calculated in terms of hadronic degrees
of freedoms in the physical side. A complete set of intermediate
physical states possessing the same quantum number as the
interpolating current are inserted into Eq.~(\ref{CorrFunc}), and
integral over $x$ is handled. After these manipulations, the
correlation function is obtained as
\begin{equation}
\Pi _{\mu \nu }^{\mathrm{Phys}}(q,T)=\frac{\langle \Psi|\eta_{\mu
}|Y_b(q)  \rangle_T \langle Y_b(q)  |\eta_{\nu }^{\dagger
}|\Psi\rangle_T }{(m_{Y_b}^{2}(T)-q^{2})}+ subtracted~terms,
\end{equation}
here $m_{Y_b}(T)$ is the temperature-dependent mass of $Y_b$
meson. Temperature dependent pole residue $\lambda_{Y_b}(T)$ is
defined in terms of matrix element as
\begin{equation}\label{Residue}
\langle \Psi|\eta_{\mu }|Y_b(q)  \rangle_T
=\lambda_{Y_b}(T)m_{Y_b}(T)~\varepsilon _{\mu},
\end{equation}
with $\varepsilon _{\mu}$ is the polarization vector of the $Y_b$
satisfying
\begin{eqnarray}\label{polarizationvec}
\varepsilon_{\mu}\varepsilon_{\nu}^{*}=-g_{\mu\nu}+\frac{q_{\mu
}q_{\nu }}{m_{Y_b}^{2}(T)}.
\end{eqnarray}
After employing polarization relations,
 the correlation function is written in terms of Lorentz structures in the form
\begin{equation}\label{CorM}
\Pi _{\mu \nu }^{\mathrm{Phys}}(q,T)=\frac{m_{Y_b}^{2}(T)\lambda_{Y_b}^{2}(T)} {%
(m_{Y_b}^{2}(T)-q^{2})} \left( -g_{\mu \nu }+\frac{q_{\mu }q_{\nu }}{m_{Y_b}^{2}(T)%
}\right) +\ldots,
\end{equation}
where dots denote the contributions coming from the continuum and higher states. To obtain the sum rules,
coefficient of any Lorentz structure can be used. In this work, coefficients of
$g_{\mu\nu}$ are chosen to construct the sum rules and the standard Borel transformation with respect to
 $q^2$ is applied to suppress the unwanted contributions. The final form of the physical side is obtained as
\begin{eqnarray}\label{CorBor}
&&\mathcal{B}(q^2)\Pi^{\mathrm{Phys}}(q,T)=m_{Y_b}^{2}(T)\lambda_{Y_b}^{2}(T)~e^{-m_{Y_b}^{2}(T)/M^{2}},
\end{eqnarray}
here $M^2$ is the Borel mass parameter. In the QCD side, $\Pi
_{\mu \nu }^{\mathrm{QCD}}(q,T)$ is calculated in terms of
quark-gluon degrees of freedom, and can be separated into two
parts over the Lorentz structures as
\begin{eqnarray}
\Pi_{\mu \nu }^{\mathrm{QCD}}(q,T)&=&\Pi _{\textit{S}}^{\mathrm{QCD}}(q^{2},T)\frac{%
q_{\mu }q_{\nu }}{q^{2}}+\Pi_{\textit{V}}^{\mathrm{QCD}}(q^{2},T)(-g_{\mu \nu }+\frac{q_{\mu }q_{\nu }}{%
q^{2}}),
\end{eqnarray}
where $\Pi _{\textit{S}}^{\mathrm{QCD}}(q^{2},T)$ and $\Pi _{\textit{V}}^{\mathrm{QCD}%
}(q^{2},T)$ are invariant functions connected with the scalar and
vector currents, respectively. In the rest framework of
$Y_b$~($\textbf{q}=0$),
$\Pi_{\textit{V}}^{\mathrm{QCD}}(q_{0}^{2},T)$ can be expressed as
a dispersion integral,
\begin{eqnarray}
\Pi_{\textit{V}}^{\mathrm{QCD}}(q_{0}^{2},T)=\int_{4(m_{b}+m_{s})^{2}}^{s_{0}(T)}\frac{\rho
^{\mathrm{QCD}}(s,T)}{(s-q_{0}^{2})}ds+...,
\end{eqnarray}
where corresponding spectral density is described as
\begin{eqnarray}
\rho^{\mathrm{QCD}}(s,T)=\frac{1}{\pi}Im\Pi_{\textit{V}}^{\mathrm{QCD}}(s,T).
\end{eqnarray}
The spectral density can be separated in terms of operator dimensions as
\begin{eqnarray}\label{Rhoall}
\rho^{\mathrm{QCD}}(s,T)&=&\rho^{\mathrm{pert.}}(s,T)+\rho^{\langle\bar{q}q\rangle}(s,T)+\rho^{\langle G^2\rangle+\langle \Theta_{00}\rangle}(s,T) \nonumber \\
&+&\rho^{\langle\bar{q}Gq\rangle}(s,T)+\rho^{\langle\bar{q}q\rangle^2}(s,T).
\end{eqnarray}
In order to obtain the expressions of these spectral density
terms, the current expression given in Eq.\ (\ref{TetraCurrent})
 is inserted into the correlation function given in Eq.\
(\ref{CorrFunc}) and then the heavy and light quark fields are
contracted, and the correlation function is written in terms of
quark propagators as
\begin{eqnarray}
\Pi _{\mu \nu }^{\mathrm{QCD}}(q,T)&=&-\frac{i}{2}\int
d^{4}xe^{iq\cdot x}\epsilon \tilde{\epsilon}\epsilon ^{\prime}\tilde{\epsilon}^{\prime}\langle%
\{\mathrm{Tr}[\gamma_{\mu}\gamma_{5}\widetilde{S}_{b}^{aa^{\prime}}(-x)\gamma_{5}\gamma_{\nu}S_{s}^{bb^{\prime}}(-x)]\nonumber \\
&\times&\mathrm{Tr}[\gamma_{5}\widetilde{S}_{s}^{dd^{\prime}}(x)\gamma_{5}S_{b}^{ee^{\prime}}(x)]
+\mathrm{Tr}[\gamma_{5}\widetilde{S}_{b}^{aa^{\prime}}(-x)\gamma_{5}S_{s}^{bb^{\prime}}(-x)\gamma_{\mu}]\nonumber \\
&\times&\mathrm{Tr}[\gamma_{5}\widetilde{S}_{s}^{dd^{\prime}}(x)\gamma_{5}\widetilde{S}_{b}^{ee^{\prime}}(x)
\gamma_{\nu}\gamma_{5}S_{b}^{bb^{\prime}}(x)]+\mathrm{Tr}[\gamma_{5}\widetilde{S}_{b}^{aa^{\prime
}}(-x)\gamma_{5}\gamma_{\nu}\nonumber \\
&\times&S_{s}^{bb^{\prime}}(-x)]\mathrm{Tr}[\gamma_{5}\widetilde{S}_{s}^{dd^{\prime}}(x)\gamma_{5}\gamma
_{\mu}S_{b}^{ee^{\prime}}(x)]+\mathrm{Tr}[\gamma_{5}\widetilde{S}_{b}^{aa^{\prime
}}(-x)\gamma _{5}\nonumber \\
&\times&S_{s}^{bb^{\prime }}(-x)] \mathrm{Tr}[ \gamma
_{5}\widetilde{S}_{s}^{dd^{\prime }}(-x)\gamma
_{5}S_{b}^{ee^{\prime }}(x)\gamma_{\nu}]\} \rangle_{T},
\end{eqnarray}
where $S_{s,b}^{ijT}(x)$ are the full quark propagators, and
$\widetilde{S}_{s,b}^{ij}(x)=CS_{s,b}^{ijT}(x)~C$ is used. The
quark propagators in vacuum are given in terms of the quark and
gluon condensates \cite{Reinders}. However, at finite
temperatures, additional operators arise due to the breaking of
Lorentz invariance by the choice of thermal rest frame. Thus, the
residual O(3) invariance brings additional operators to the quark
propagator at finite temperature. The expected behavior of the
thermal averages of these new operators is opposite of those of
the Lorentz invariant old ones \cite{Mallik:1997pq}. The thermal
heavy-quark propagator in coordinate space can be expressed as
\begin{eqnarray}\label{HeavyProp}
S_{b}^{ij}(x)&=&i\int \frac{d^{4}k}{(2\pi )^{4}}e^{-ik\cdot x}\Bigg[ \frac{%
\delta _{ij}\Big( {\!\not\!{k}}+m_{b}\Big)
}{k^{2}-m_{b}^{2}}-\frac{gG_{ij}^{\alpha \beta }}{4}\frac{\sigma _{\alpha \beta }\Big( {%
\!\not\!{k}}+m_{b}\Big) +\Big(
{\!\not\!{k}}+m_{b}\Big)\sigma_{\alpha
\beta }}{(k^{2}-m_{b}^{2})^{2}}\nonumber \\
&+&\frac{g^{2}}{12}G_{\alpha \beta }^{A}G_{A}^{\alpha \beta
}\delta_{ij}m_{b}\frac{k^{2}+m_{b}{\!\not\!{k}}}{(k^{2}-m_{b}^{2})^{4}}+\ldots\Bigg],
\end{eqnarray}
where $G_{A}^{\alpha \beta}$ is the external gluon field, $A=
1,2,...8$,~$\lambda_{ij}^{A}$ are the Gell-Mann matrices,
$t_{ij}^{A}=\lambda_{ij}^{A}/2$. The thermal light-quark
propagator is chosen as
\begin{eqnarray} \label{LQprop}
S_{s}^{ij}(x)
&=&i\frac{{\!\not\!{x}}}{2\pi^{2}x^{4}}\delta_{ij}-\frac{m_s}{4\pi^{2}x^2}\delta_{ij}
-\frac{\langle\bar{s}s\rangle}{12}\delta_{ij}-\frac{x^2}{192}m_{0}^{2}\langle
\bar{s}s\rangle \Big[1-i\frac{m_{s}}{6}{\!\not\!{x}}
\Big]\delta_{ij}\nonumber\\
&+&\frac{i}{3}\Big[{\!\not\!{x}}\Big(\frac{m_{s}}{16}\langle
\bar{s}s\rangle -\frac{1}{12}\langle u\Theta ^{f}u\rangle \Big)
+\frac{1}{3}\Big(u\cdot x\Big){\!\not\!{u}} \langle u\Theta^{f}u\rangle %
\Big]\delta _{ij} \nonumber\\
&-&\frac{ig_{s}\lambda_{ij}^{A}}{32\pi ^{2}x^{2}}G_{A}^{\mu \nu }
\Big({\!\not\!{x}} \sigma _{\mu \nu }+\sigma _{\mu \nu
}{\!\not\!{x}}\Big),
\end{eqnarray}
where $m_{s}$ implies the strange quark mass, $u_{\mu }$ is the
four-velocity of the heat bath, $\langle \bar{q}q\rangle $ is the
temperature-dependent light quark condensate and $\Theta _{\mu \nu
}^{f}$ is the fermionic part of the energy momentum tensor.
Furthermore, the gluon condensate related to the gluonic part of
the energy-momentum tensor $\Theta_{\alpha \beta }^{g}$ is defined
via relation~\cite{Mallik:1997pq}:
\begin{eqnarray}\label{TrGG}
< Tr^c G_{\alpha \beta }G_{\lambda \sigma}>_T &=&
(g_{\alpha\lambda} g_{\beta\sigma}- g_{\alpha\sigma}
g_{\beta\lambda})A \nonumber\\ &-& (u_{\alpha} u_{\lambda}
g_{\beta\sigma}- u_{\alpha} u_{\sigma} g_{\beta\lambda} -
u_{\beta} u_{\lambda} g_{\alpha\sigma} + u_{\beta} u_{\sigma}
g_{\alpha\lambda})B,
\end{eqnarray}
where $A$ and $B$ coefficients are
\begin{eqnarray}\label{AB}
A &=& \frac{1}{24} < G^a_{\alpha \beta }G^{a \alpha \beta}>_T +
\frac{1}{6} < u^\alpha \Theta^{g}_{\alpha \beta }
u^\beta>_T,\nonumber\\
B &=& \frac{1}{3} < u^\alpha \Theta^{g}_{\alpha \beta }
u^\beta>_T.
\end{eqnarray}
In order to remove contributions originating from higher states,
the standard Borel transformation with respect to $q_0^{2}$ is
applied in the QCD side as well. By equating the coefficients of
the selected structure $g_{\mu \nu}$ in both physical and QCD
sides, and by employing the quark hadron duality ansatz up to a
temperature dependent continuum threshold $s_0(T)$, the final sum
rules for $Y_b$ are derived as
\begin{equation}\label{ResidueSR}
m_{Y_b}^{2}(T)\lambda_{Y_b}^{2}(T)~e^{-m_{Y_b}^{2}(T)/M^{2}}=
\int_{4(m_{b}+m_{s})^{2}}^{s_{0}(T)}ds~\rho
^{\mathrm{QCD}}(s,T)~e^{-s/M^{2}}.
\end{equation}
To find the mass via TQCDSR, one should expel the hadronic
coupling constant from the sum rules. It is commonly done by
dividing the derivative of
 the sum rule given in Eq.\ (\ref{ResidueSR}) with respect to $(-M^{-2})$
to itself. Following these steps, the temperature dependent mass is obtained as
\begin{equation}\label{massSR}
m_{Y_b}^{2}(T)=\frac{\int_{4(m_{b}+m_{s})^{2}}^{s_{0}(T)}ds~s~\rho ^{\mathrm{QCD}%
}(s,T)~e^{-s/M^{2}}}{\int_{4(m_{b}+m_{s})^{2}}^{s_{0}(T)}ds~\rho^{\mathrm{QCD}}
(s,T)~e^{-s/M^{2}}},
\end{equation}
where the thermal continuum threshold $s_0(T)$ is related to
continuum threshold $s_0$ at vacuum via
relation\cite{Dominguez:2009mk,Veliev:2010gb}
\begin{eqnarray}  \label{G2TLattice}
s_0(T)=s_0\left[1-\left(\frac{T}{T_c}\right)^8\right]
+4(m_b+m_s)^2\left(\frac{T}{T_c}\right)^8.
\end{eqnarray}
For compactness, the explicit forms of spectral densities are
presented in Appendix.

\section{Numerical Analysis}\label{sec:NumAnal}

\begin{figure}[htbp]
\begin{center}
\includegraphics[width=0.47\textwidth]{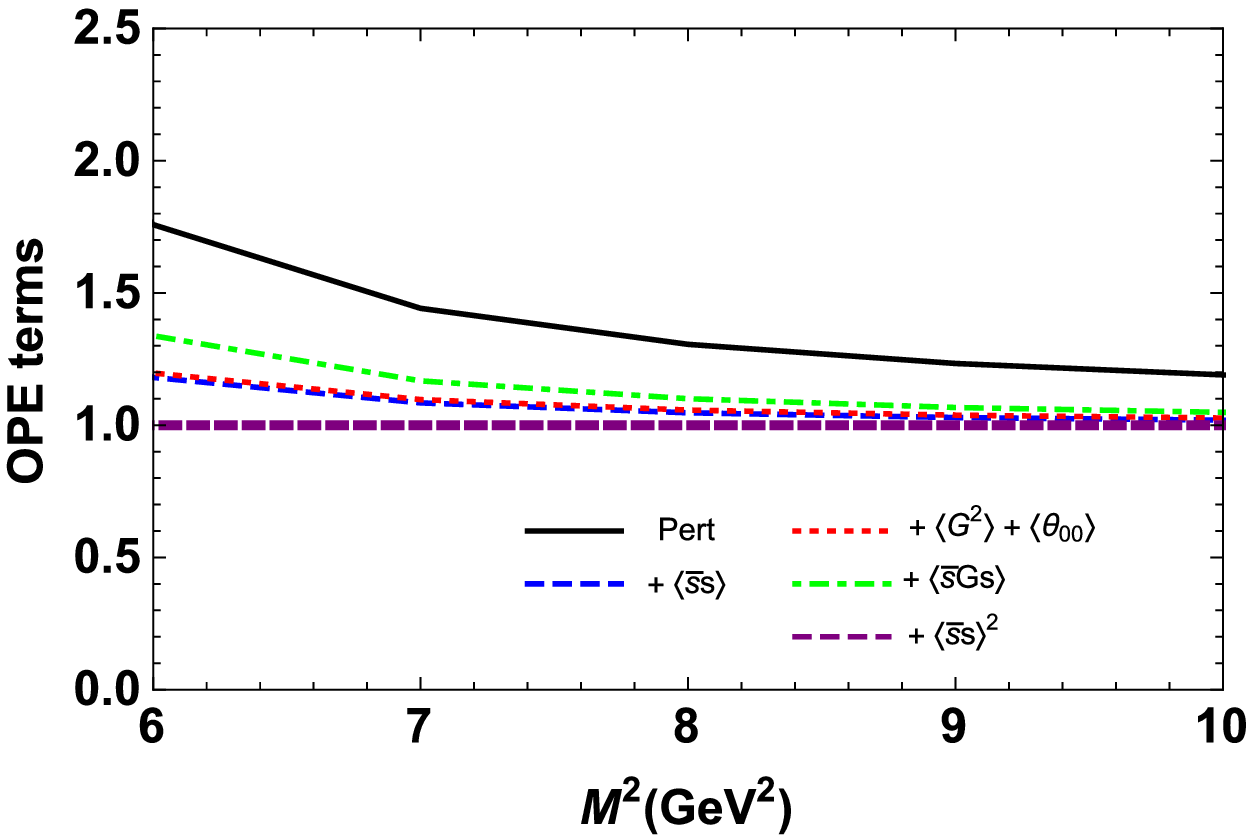}\hskip0.5cm\includegraphics[width=0.47\textwidth]{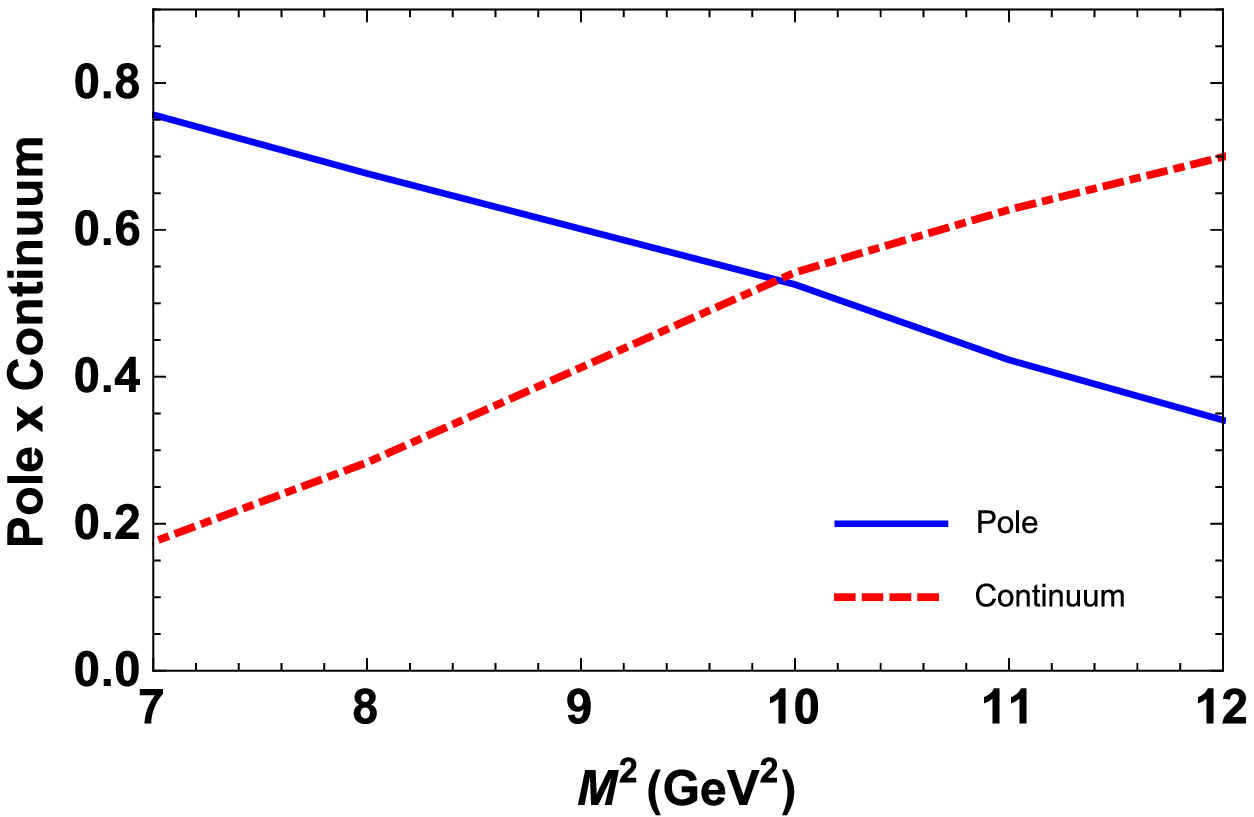}
\end{center}
\caption{The OPE convergence of the sum rules: The ratio of the
sum of the contributions up to specified dimension to the total
contribution is plotted with respect to ~$M^2$ at $s_0=134
~\mathrm{GeV^2}, T=0$ ~(left)\label{fig:Ope}. Pole dominance of
the sum rules: relative contributions of the pole~(blue) and
continuum~(red-dashed) versus the Borel parameter $M^2$ at
$s_0=134~\mathrm{GeV^2}, T=0$ (right).\label{fig:Pole}}
\end{figure}

\begin{figure}[htbp]
\begin{center}
\includegraphics[width=0.47\textwidth]{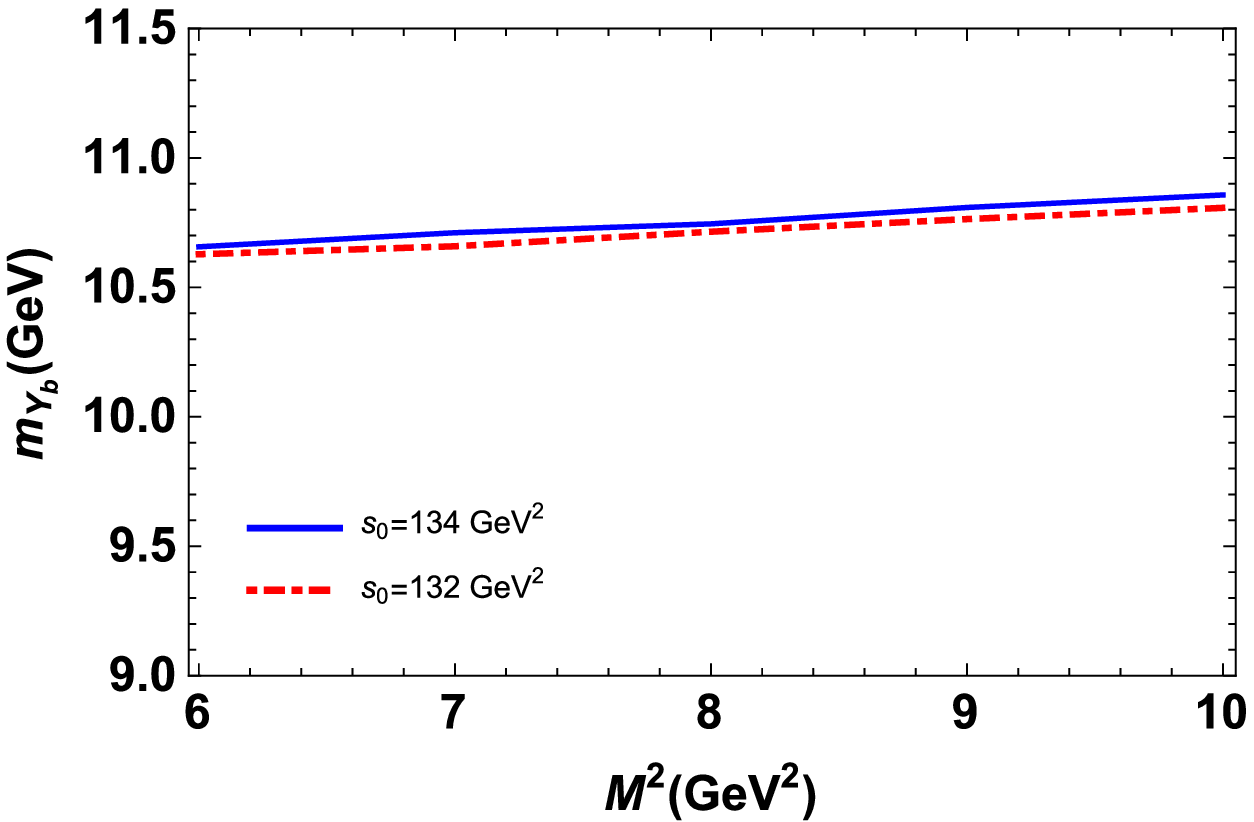}\hskip0.5cm\includegraphics[width=0.485\textwidth]{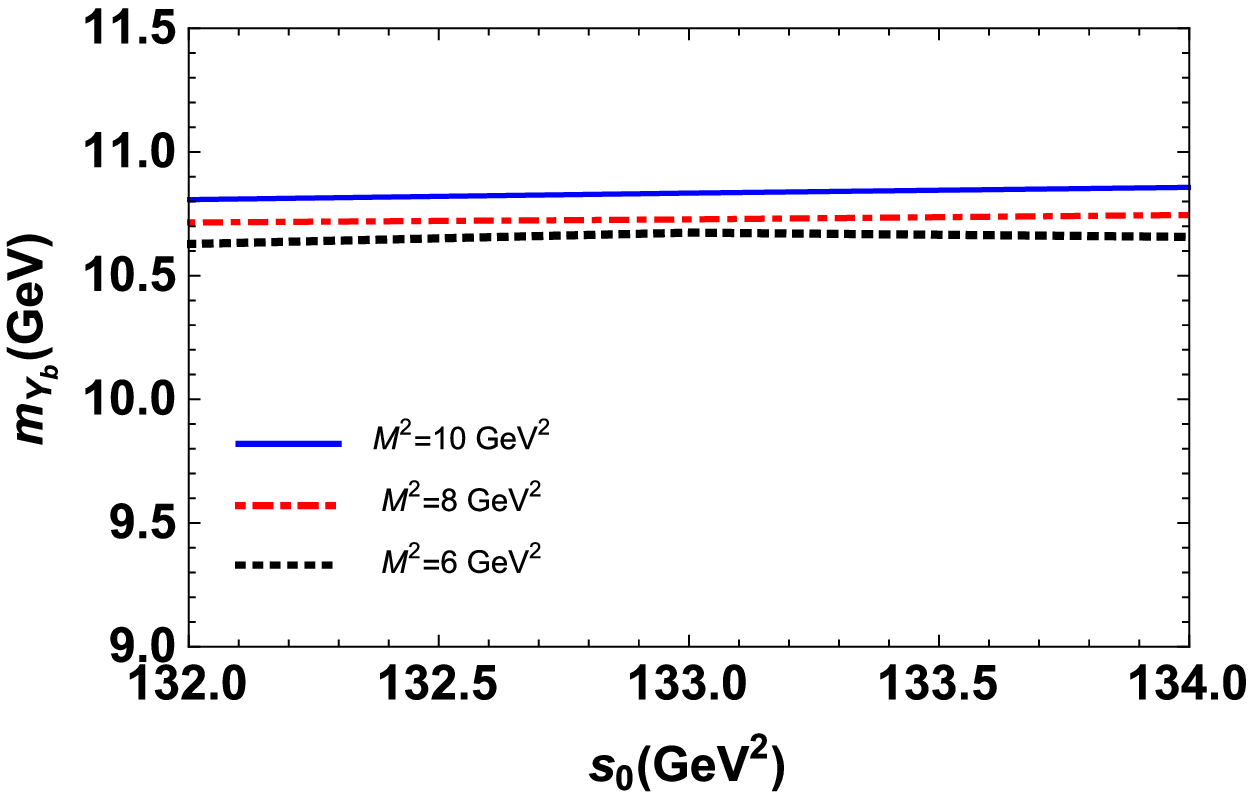}
\end{center}
\caption{Mass of $Y_b$ as a function of $M^2$ (left) and $s_0$
(right). \label{fig:Results}}
\end{figure}

In this section, numerical analysis to obtain the values of the mass and the pole
residue of $Y_b$ state at vacuum and also $T\neq0$ cases is presented.
By following the analysis, one can see the hot medium effects
on the hadronic parameters of the $Y_b$ state. During the
calculations, input parameters given in Table \ref{tab:inputPar} are used.
\begin{table}[htbp]
\begin{center}
\caption{\label{tab:inputPar}Input
parameters~\cite{Tanabashi,Dosch:1988vv,Belyaev:1982cd,Ioffe:2005ym}}
\item[]\begin{tabular}{@{}l} \br
$ m_s=(0.13\pm0.03)~\mathrm{MeV}$  \\
 $ m_b=(4.24\pm0.05)~ \mathrm{GeV}$  \\
 $ m^2_0=(0.8\pm0.2)\mathrm{GeV}^2$\\
 $\langle s\bar{s}\rangle=-0.8\times(0.24\pm0.01)^3~\mathrm{GeV}^3$ \\
$\langle0|\frac{1}{\pi}\alpha_sG^2|0\rangle=(0.012~\mathrm{GeV}^4)$ \\
\br
\end{tabular}
\end{center}
\end{table}
In addition to these input parameters,
temperature-dependent quark and gluon condensates, and the energy
density expressions are necessary. The thermal quark condensate is chosen as
\begin{equation}\label{eq::qqT}
\langle \bar{q}q\rangle =\frac{\langle 0|\bar{q}q|0\rangle }{%
1+exp~\Bigg({18.10042\big(1.84692\big[\frac{1}{\mathrm{\mathrm{GeV}}^{2}}\big]T^{2}+4.99216\big[\frac{1}{%
\mathrm{\mathrm{GeV}}}\big]T-1\big)}\Bigg)},
\end{equation}
where $\langle 0|\bar{q}q|0\rangle$ is the light quark condensate
at vacuum, and which is credible up to a critical temperature
$T_{c}=190~\mathrm{\mathrm{MeV}}$. The expression given in Eq.
(\ref{eq::qqT}) is obtained in Refs. \cite{Azizi:2016ddw,Ayala}
from the Lattice QCD results given in Refs. \cite{Bazavov,Cheng1}.
The temperature-dependent gluon condensate is parameterized via
\cite{Azizi:2016ddw,Ayala2}
\begin{eqnarray}\label{G2TLattice}
\langle G^{2}\rangle  =\langle 0|G^{2}|0\rangle \Bigg[1-1.65\Bigg(\frac{T}{%
T_{c}}\Bigg)^{8.735}+0.04967\Bigg(\frac{T}{T_{c}}\Bigg)^{0.7211}\Bigg],
\end{eqnarray}
where $\langle 0|G^{2}|0\rangle $ is the gluon condensate in
vacuum state and $G^2 = G^A_{\alpha\beta}G_A^{\alpha\beta}$.
Additionally, for the gluonic and fermionic parts of the energy
density, the following parametrization is used
~\cite{Azizi:2016ddw}
\begin{eqnarray}\label{tetamumu}
\langle \Theta _{00}^{g}\rangle &=&\langle \Theta^f
_{00}\rangle=T^{4}exp~{\Big(113.867\Big[\frac{1}{\mathrm{GeV}^{2}}\Big]T^{2}-12.190\Big[\frac{1}{\mathrm{GeV}}\Big]T
\Big)} \nonumber \\
&-& 10.141\Big[\frac{1}{\mathrm{GeV}}\Big]T^{5},
\end{eqnarray}
which is extracted from the Lattice QCD data in
Ref.~\cite{Cheng:2007jq}.
In order to get reliable results, obtained sum rules should be
tested at vacuum, and the working regions of the parameters $s_0$
and $M^2$ should be determined. Within the working regions of
$s_0$ and $M^2$, convergence of OPE and dominance of pole
contributions should be assured. In addition, the obtained
physical results should be independent of small variations of
these parameters. Convergence of the OPE is tested by the
following criterion. The contribution of the highest order
operator in the OPE should be very small compared to the total
contribution. In Fig.~\ref{fig:Ope}, the ratio of the sum of the
terms up to the specified dimension to the total contribution is
plotted to test the OPE convergence. It is seen that all higher
order terms contribute less than the perturbative part for
$M^2\geq 6$ GeV$^2$. On the other hand, dominance of the pole
contribution is tested as follows. The contribution coming from
the pole of the ground state should be greater than the
contribution of the continuum. In this work, the aforementioned
ratio is
\begin{equation}
\mathrm{PC}=\frac{\Pi (M_{\mathrm{max}}^{2},\ s_{0})}{\Pi (M_{\mathrm{max}%
}^{2},\ \infty )}>0.50,  \label{eq:Rest1}
\end{equation}
when $M^2 \leq 10$ GeV$^2$ as can be seen in Fig. \ref{fig:Ope}.
After checking these criteria, the working regions of the
parameters $M^2$ and $s_0$ are determined as
\begin{eqnarray*}
6~\mathrm{GeV}^2\leq M^2 \leq 10~\mathrm{GeV}^2~~~ ;~~~~~~
132~\mathrm{GeV}^2\leq s_0 \leq 134~\mathrm{GeV}^2,
\end{eqnarray*}
which is also consistent with $s_0\simeq (m_H+0.5$ GeV$)^2$ norm
\cite{Albuquerque:2011ix}. Within these working regions, the
variations of the mass of $Y_b$ with respect to $M^2$ and $s_0$
are plotted in Figure~\ref{fig:Results}. It is seen that the mass
is stable with respect to variations of $M^2$ and $s_0$. In Table~
\ref{tab:Results1}, the mass obtained in this work is presented
together with the ones appearing in literature, and it is
estimated in good agreement with other theoretical estimates and
as well as experimental results
\cite{Tanabashi,Zhang:2010mv,Albuquerque:2011ix}. Thus, the broad
resonance $Y_b$ can be described by the tetraquark current given
in Eq.~(\ref{TetraCurrent}).

\begin{table}[htbp]
\caption{Results obtained in this work for the mass of $Y_b$ at
$T=0$, in comparison with literature.\label{tab:Results1}}
\begin{center}
\item[]\begin{tabular}{@{}lll} \br
    & $m_{Y_b}(\mathrm{MeV})$       \\
 \mr
Present Work   & $10735^{+122}_{-107}$                  \\
\hline
 Experiment   & $10889.9^{+3.2}_{-2.6}$ \cite{Tanabashi}      \\
\hline
QCDSR                & $10880 \pm 130$  \cite{Zhang:2010mv}          \\
 & $10910 \pm 70$    \cite{Albuquerque:2011ix}   \\
\br
\end{tabular}
\end{center}
\end{table}
\begin{figure}[htbp]
\begin{center}
\includegraphics[width=0.47\textwidth]{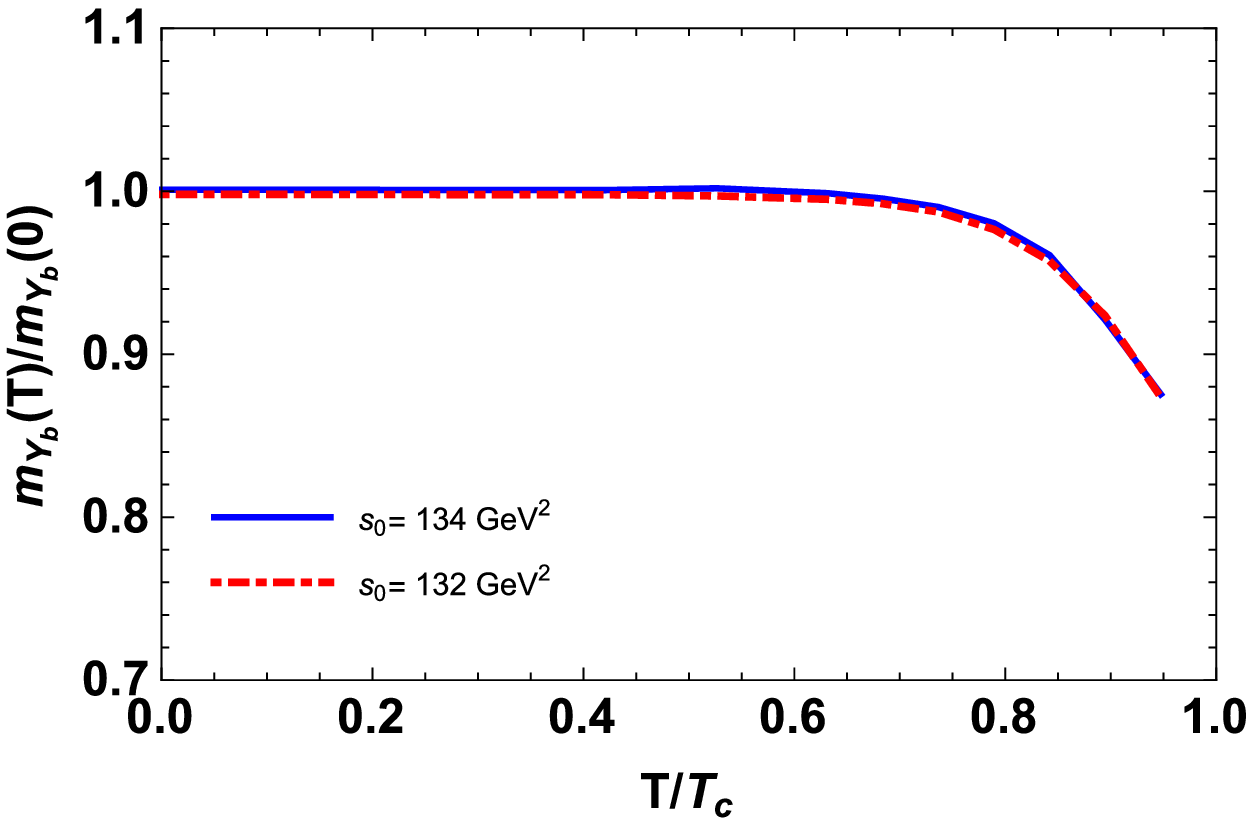}\hskip0.5cm
\includegraphics[width=0.47\textwidth]{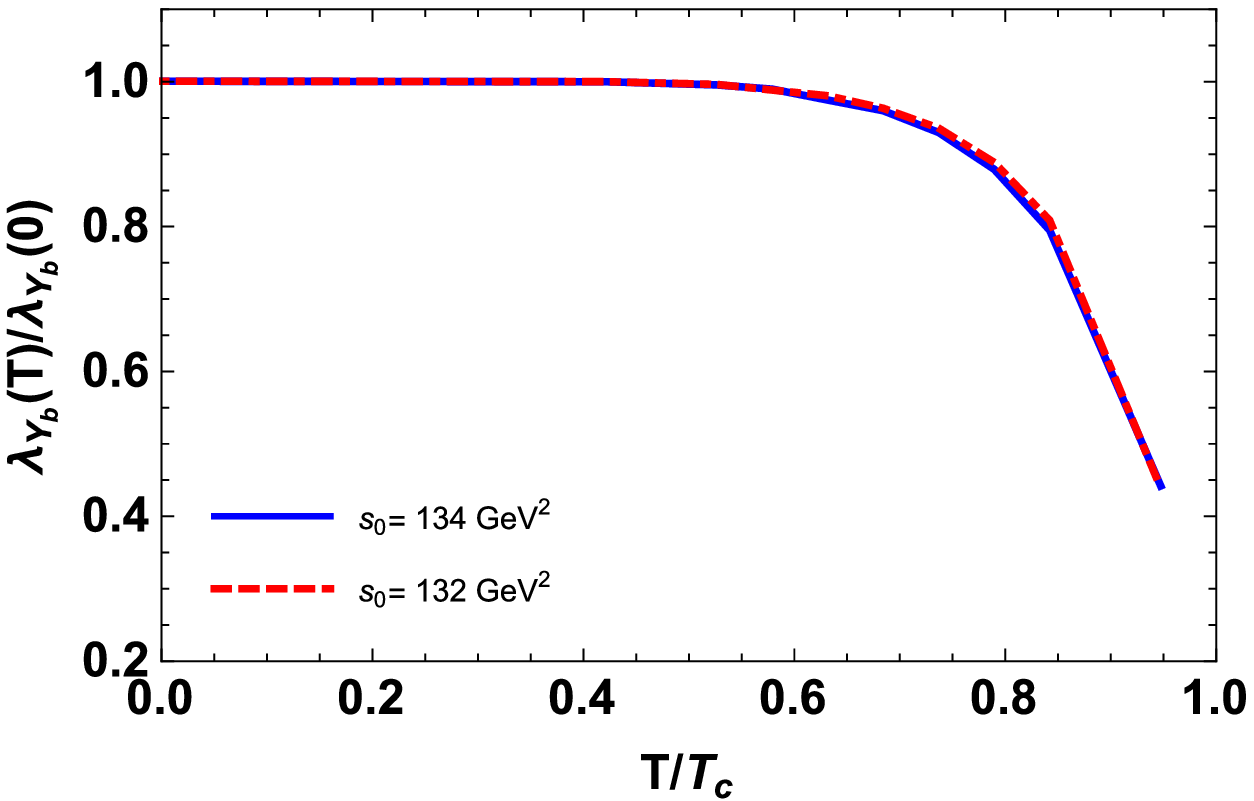}
\end{center}
\caption{The mass (left) and pole residue (right) of $Y_b$ as a
function of temperature.\label{fig:temp}}
\end{figure}

After testing the sum rules at $T=0$, and comparing the obtained
results, the temperature dependence of the mass and the pole
residue of $Y_b$ is plotted in Figure~\ref{fig:temp}. Thermal
behavior of mass and pole residue of $Y_b$ are monotonous until $T
\cong 0.12~\mathrm{\mathrm{GeV}}$. However, after this point, they
begin to decrease promptly with increasing temperature. At the
vicinity of the critical (or so called deconfinement) temperature,
the mass reaches nearly $87\%$ of its vacuum value. On the other
hand, the pole residue is decreased to $44\%$ of its value at
vacuum. Even though the thermal dependencies indicate that the
mass and the pole residue of $Y_b$ might dissolve at $T_c$, its
temperature dependent decay behavior should also be studied to
conclude on the stability of the state. Since the decay properties
also depend on temperature, and while the mass and pole residue
diminish, the decay width might increase with increasing
temperature \cite{Azizi:2010zza}, decay widths at finite
temperature should also be investigated. However the current
status of $Y_b$ resonance is very complicated, since it is very
close to $\Upsilon(5S)$. Thus studying its decays requires
establishment of a good model in the hidden-beauty sector.

\section{Concluding Remarks}\label{sec:Result}

In this work, we revisited the hidden beauty exotic state $Y_b$,
and studied its properties at vacuum and finite temperatures. To
describe the hot medium effects to the hadronic parameters of the
resonance $Y_b$, TQCDSR method is used considering contributions
of condensates up to dimension six. Our results are in reasonable
agreement with the available experimental data and other vacuum
QCD sum rules works in the literature. Numerical findings show
that $Y_b$ can be well described by a scalar-vector tetraquark
current. In addition, we observed that until temperature
$T\simeq120~\mathrm{MeV}$ the mass and pole residue values stay
nearly same with QCD vacuum values. However after this point
increasing temperature behaviors of the mass and the pole residue
is changed remarkably, and they reduced approximately to $87\%$
and $44\%$ of their vacuum values, respectively. These findings
are in good agreement with the temperature behavior of
conventional hadrons, and also with exotic states. In the
literature, remarkable drop in the values of the mass and the pole
residue in hot medium was regarded as the signal of the QGP, which
is called as the fifth state of matter, phase transition. Also,
studying of $Y_b$ state in hot medium can be a useful tool to
analyze results of the heavy-ion collision experiments. We hope
that precise spectroscopic measurements in the exotic bottomonium
sector can be done at Super-B factories, and this might provide
conclusive answers on the nature and thermal behaviors of the
exotic states.

\appendix

\section{Thermal spectral density $\rho^{\mathrm{QCD}}(s, T)$
for $Y_b$ state}\label{sec:App}

In this appendix, the explicit forms of the spectral densities
obtained in this work are presented. The expressions for
$\rho^{\mathrm{pert.}}(s)$ and $\rho^{\mathrm{nonpert.}}(s,T)$ are
shown below as integrals over the Feynman parameters $z$ and $w$,
where $\theta$ is the step function,
\begin{eqnarray}\label{RhoPert}
\rho^{\mathrm{pert.}}(s)&=&\frac{1}{3072\pi^6}\int_{0}^{1}dz\int_{0}^{1-z}dw \nonumber \\
&\times& \frac{1}{\kappa^8 \xi^2} \Bigg\{\Big[-\kappa m_b^2 (z+w)+s z w \xi\Big]^2 \Big[\kappa^2 m_b^4 z w (z+w)[w^2+(-1+w)\nonumber \\
&\times& w+z (-1+4 w)]-2 \kappa m_b^2 \Big[6 m_s^2 \Phi^2 [7(-1+z) z (-7+8 z)\nonumber \\
&\times& w +7 w^2]+s z^2 w^2 (12 (-1+z) z+(-12+25z) w \nonumber \\
&+&12 w^2)\Big]\xi+s z w (12 m_s^2 \Phi^2 +35 s z^2
w^2)\xi^3\Big]\Bigg\}~\theta[L(s,w,z)],
\end{eqnarray}
\begin{eqnarray}\label{RhoDim3}
\rho^{\langle s\bar{s}\rangle}(s,T)&=&\frac{\langle\bar{s}s\rangle}{128\pi^4}\int_{0}^{1}dz \int_{0}^{1-z}dw%
 \frac{1}{\kappa^6}\Bigg\{\Big[\kappa^3 m_b^5 w (z + w)^2+\kappa^2 m_b^4 m_s (z + w) \nonumber \\
&\times&[19 z^4 + 19 (-1 + w)^2 w^2+ 2 z (-1 + w) w (-19 + 25 w)+z^3 \nonumber \\
&\times&(-38 + 50 w)+ z^2 [19 + w(-88 + 81 w)]\Big] -\kappa^2 m_b^3 z (z + w)\nonumber \\
&\times& (m_s^2 \Phi + 2 s w^2)\xi -\kappa m_b^2 m_s s z w \Big[22 z^4 + 22 (-1 + w)^2 w^2 +z^3 \nonumber \\
&\times& (-44 + 111 w) +z (-1 + w) w (-44 + 111 w) + z^2 [22 +w \nonumber \\
&\times&(-155 + 197 w)]\Big] \xi+ \kappa m_b s z^2 w (m_s^2 \Phi + s w^2) \xi^2 + 3 m_s s^2 z^2 \nonumber \\
&\times& w^2 \Big(z^2 + (-1 + w) w + z (-1 + 21 w)\Big)
\xi^3\Big]\Bigg\}~\theta[L(s,w,z)],
\end{eqnarray}
\begin{eqnarray}\label{RhoDim4}
\rho^{\langle G^2 \rangle+\langle \Theta_{00} \rangle}(s,T)&=&\frac{1}{4608 \pi^2 }\int_{0}^{1}dz\int_{0}^{1-z}dw \nonumber \\
&\times&\frac{1}{\kappa^6\xi^2}\Bigg\{192 \pi^2 \langle  \Theta^f_{00} \rangle zw\xi^2\Bigg[m_b^4 \Phi^2 (z + w) [3 (-1 + z) z + (-3 + 5 z) \nonumber \\
&\times& w + 3 w^2] + m_b^2 \Phi s z w [-27 (-1 + z) z + 27 w -53zw - 27 w^2]  \nonumber \\
&\times& \xi+ 30 s^2 z^2 w^2 \xi^3\Bigg] - g_s^2 \langle \Theta^g_{00}\rangle \Big[3m_b^4 \Phi^2 z w (z + w) [2 (-1 + z)^2 z^2 \nonumber\\
&+& (-1 + z) z (-4 + 3z) w + (-2 + z) (-1 + 3 z) w^2 +(-4 + 3 z)\nonumber\\
&\times& w^3 + 2 w^4] - m_b^2 \Phi s z^2 w^2 [24 (-1 + z)^2 z^2 +(-1 + z) z (-48+85  \nonumber\\
&\times& z) w + [24 + z (-133 + 121 z)] w^2 + (-48 + 85 z) w^3+24w^4] \xi \nonumber\\
&-& 12 m_b^3 m_s \Phi^3 (z +w)^3 \xi^2 + 12 m_b m_s \Phi^2 s z z (z^2 + 10 z w + w^2)\xi^3+30 \nonumber\\
&\times& s^2z^3w^3(z+w)\xi^4\Big]+\langle \frac{\alpha_s G^2}{\pi}\rangle \pi^2\Bigg[\kappa m_b^2\Big\{-6 m_s^2 \Phi^2 \big[5 (-1+z)z^3+ z^3 \nonumber\\
&\times& w +(-5 + z) w^3 + 5 w^4\big] + s z^2 w^2 [2 (-1 + z)z^2 (-18+11 z)+z \nonumber\\
&\times& [72 + z (-221 + 133 z)] w + [36 + z (-221 + 231 z)] w^2(-58+133  \nonumber\\
&\times& z)w^3 + 22 w^4]\Big\}\xi - 36 \kappa^3 m_b^3 m_s (z+ w) (z^2 - 6 z w+w^2) \xi^2 - 60 s^2\nonumber\\
&\times&  z^3 w^3 (z +w) \xi^4 + \xi^2 m_b z w [m_b^3 (z + w) (4 (-1 + z)z^2(-3 + 4 z)\nonumber\\
&+& 6 z [4 + z (-11 + 8 z)] w + [12 + 11 z(-6 + 5 z)] w^24(-7+12z) \nonumber\\
&\times&w^3+16w^4)+12m_{s}s(3z^2-26zw+3w^2)\xi^3]\Bigg]\Bigg\}~\theta[L(s,w,z)]
\end{eqnarray}
\begin{eqnarray}\label{RhoDim5part1and2}
\rho^{\langle \bar{s}G s\rangle}(s,T)&=&\frac{m_s m_0^2\langle\overline{s}s\rangle}{64\pi^4}\Bigg\{\int_{0}^{1}dz\Big\{3m_b^2+s(z-1)z~\theta[L'(s,z)]\Big\}\nonumber \\
&+&\int_{0}^{1}dz \int_{0}^{1-z}dw \frac{1}{3\kappa^5}\Big\{z w \xi\Big[\kappa m _b^2[5z^2+5(w-1)w+z(11w-5)] \nonumber \\
&-&16szw\xi^2\Big]~\theta[L(s,w,z)]\Big\}\Bigg\},
\end{eqnarray}
\begin{eqnarray}\label{RhoDim6part1and2}
\rho^{\langle \overline{s}s
\rangle^2}(s,T)&=&\frac{\langle\overline{s}s\rangle^2}{5184\pi^4}\Bigg\{\int_{0}^{1}dz\Big[648
m_b^2\pi^2 + g_s^2 m_b m_s z + 54 \pi^2 (5 m_s^2 + 4 s)(z-1) z\Big] \nonumber \\
&\times&\theta[L'(s,z)]+g_s^2\int_{0}^{1}dz\int_{0}^{1-z}\frac{dw
zw\xi}{\kappa^5}\Big[3\kappa m_b^2 [7z^2+7(w-1)w \nonumber \\
&+& z(-7+15w)] -64 szw \xi^2\Big]~\theta[L(s,w,z)]\Bigg\},
\end{eqnarray}
where explicit expressions of the functions $L(s,w,z)$ and
$L'(s,z)$ are
\begin{eqnarray}
L[(s,w,z)&=& \frac{(-1 + w) \Big[(-1 + w) w^2 + 2 (-1 + w) w z +
(-1 +2w) z^2 - s w z \xi + z^3 m_b^2\Big]}{\kappa^2},\nonumber \\
L'(s,z)&=& s z (1-z)-m_b^2.
\end{eqnarray}
The below definitions are used for simplicity:
\begin{eqnarray}
\kappa &=&z^2 + z (w-1) + (w-1) w,  \nonumber \\
\Phi &=&(z-1) w+ (z-1) w + w^2,  \nonumber \\
\xi&=& z + w-1.
\end{eqnarray}

\section*{Acknowledgment}

\label{sec:e}

J. Y. S\"{u}ng\"{u}, A. T\"{u}rkan and E. Veli Veliev thank to
Kocaeli University for the partial financial support through the
grant BAP 2018/082.  H. Da\u{g} acknowledges support through the
Scientific and Technological Research Council of Turkey (TUBITAK)
BIDEP-2219 grant.

\end{document}